\documentclass[12pt,epsf]{article}
\usepackage{graphicx}
\begin{document}

\def\a{\alpha}
\def\b{\beta}
\def\c{\varepsilon}
\def\d{\delta}
\def\e{\epsilon}
\def\f{\phi}
\def\g{\gamma}
\def\h{\theta}
\def\k{\kappa}
\def\l{\lambda}
\def\m{\mu}
\def\n{\nu}
\def\p{\psi}
\def\q{\partial}
\def\r{\rho}
\def\s{\sigma}
\def\t{\tau}
\def\u{\upsilon}
\def\v{\varphi}
\def\w{\omega}
\def\x{\xi}
\def\y{\eta}
\def\z{\zeta}
\def\D{\Delta}
\def\G{\Gamma}
\def\H{\Theta}
\def\L{\Lambda}
\def\F{\Phi}
\def\P{\Psi}
\def\S{\Sigma}

\def\o{\over}
\newcommand{\gsim}{ \mathop{}_{\textstyle \sim}^{\textstyle >} }
\newcommand{\lsim}{ \mathop{}_{\textstyle \sim}^{\textstyle <} }
\newcommand{\vev}[1]{ \left\langle {#1} \right\rangle }
\newcommand{\bra}[1]{ \langle {#1} | }
\newcommand{\ket}[1]{ | {#1} \rangle }
\newcommand{\EV}{ {\rm eV} }
\newcommand{\KEV}{ {\rm keV} }
\newcommand{\MEV}{ {\rm MeV} }
\newcommand{\GEV}{ {\rm GeV} }
\newcommand{\TEV}{ {\rm TeV} }
\def\diag{\mathop{\rm diag}\nolimits}
\def\Spin{\mathop{\rm Spin}}
\def\SO{\mathop{\rm SO}}
\def\O{\mathop{\rm O}}
\def\SU{\mathop{\rm SU}}
\def\U{\mathop{\rm U}}
\def\Sp{\mathop{\rm Sp}}
\def\SL{\mathop{\rm SL}}
\def\tr{\mathop{\rm tr}}

\def\IJMP{Int.?J.?Mod.?Phys. }
\def\MPL{Mod.?Phys.?Lett. }
\def\NP{Nucl.?Phys. }
\def\PL{Phys.?Lett. }
\def\PR{Phys.?Rev. }
\def\PRL{Phys.?Rev.?Lett. }
\def\PTP{Prog.?Theor.?Phys. }
\def\ZP{Z.?Phys. }

\newcommand{\bear}{\begin{array}}  \newcommand{\eear}{\end{array}}
\newcommand{\bea}{\begin{eqnarray}}  \newcommand{\eea}{\end{eqnarray}}
\newcommand{\beq}{\begin{equation}}  \newcommand{\eeq}{\end{equation}}
\newcommand{\bef}{\begin{figure}}  \newcommand{\eef}{\end{figure}}
\newcommand{\bec}{\begin{center}}  \newcommand{\eec}{\end{center}}
\newcommand{\non}{\nonumber}  \newcommand{\eqn}[1]{\beq {#1}\eeq}
\newcommand{\la}{\left\langle} \newcommand{\ra}{\right\rangle}

\def\SEC#1{Sec.?\ref{#1}}
\def\FIG#1{Fig.?\ref{#1}}
\def\EQ#1{Eq.?(\ref{#1})}
\def\EQS#1{Eqs.?(\ref{#1})}
\def\lrf#1#2{ \left(\frac{#1}{#2}\right)}
\def\lrfp#1#2#3{ \left(\frac{#1}{#2}\right)^{#3}}
\def\GEV#1{10^{#1}{\rm\,GeV}}
\def\MEV#1{10^{#1}{\rm\,MeV}}
\def\KEV#1{10^{#1}{\rm\,keV}}
\def\REF#1{(\ref{#1})}
\def\lrf#1#2{ \left(\frac{#1}{#2}\right)}
\def\lrfp#1#2#3{ \left(\frac{#1}{#2}\right)^{#3}}


\baselineskip 0.7cm

\begin{titlepage}

\begin{flushright}
IPMU08-0082
\end{flushright}

\vskip 1.35cm
\begin{center}
{\large \bf
Gravity Mediation of Supersymmetry Breaking \\ with Dynamical Metastability
}
\vskip 1.2cm
Izawa K.-I.$^{1,2}$, Fuminobu Takahashi$^{2}$, T.T.~Yanagida$^{2,3}$, and Kazuya Yonekura$^{3}$
\vskip 0.4cm
{\it
${}^1$Yukawa Institute for Theoretical Physics, Kyoto University,\\
Kyoto 606-8502, Japan\\
${}^2$Institute for the Physics and Mathematics of the Universe,
University of Tokyo,\\ Chiba 277-8568, Japan\\
${}^3$Department of Physics, University of Tokyo,\\
  Tokyo 113-0033, Japan}

\vskip 1.5cm

\abstract{
We argue that the Polonyi problem can be avoided
when our supersymmetry-breaking vacuum
is surrounded by many supersymmetric vacua.
We construct a dynamical class of supersymmetry-breaking 
models to demonstrate our point.
These models naturally predict a small
deviation from the standard big-bang nucleosynthesis.
}
\end{center}
\end{titlepage}

\setcounter{page}{2}

\section{Introduction}
\label{sec:1}

Theoretical landscape such as in string/M theory
is expected to possess an enormous number
of metastable vacua in addition to supersymmetric (SUSY) ones.
With such vacua sufficiently long-lived in hand,
we are led to suspect that our SUSY-breaking state is indeed metastable.

The possibility we pursue in this paper is that
the Polonyi problem in gravity mediation of SUSY breaking
can be avoided when our SUSY-breaking vacuum is surrounded
by many SUSY vacua. Then, the initial value
of the Polonyi field $S$ should be chosen not too far away from our
vacuum, since otherwise, the $S$ would roll down to the SUSY vacua.

We provide a class of dynamical models to demonstrate the above point,
which may be of some interest in its own right.

\section{Dynamical Models}
\label{sec:2}

Let us consider supersymmetric QCD-like theory of the gauge group $SU(N_C)$
and $N_F$ quark flavors $Q_i$, ${\tilde Q}_i$ ($i=1, \cdots, N_F$;
the gauge indices omitted)
with a superpotential
\beq
W = \l S \sum_i Q_i {\tilde Q}_i,
\eeq
where $S$ is a singlet chiral superfield
and $\l$, $N_C$, $N_F$ are positive constants of order one.

With the other fields integrated out, the effective superpotential
of $S$ is given by
\beq
W_{{\rm eff}} = { \L^3 \o 16\pi^2} \left( {\l S \o \L} \right)^{N_F \o N_C},
\eeq
where $\L$ denotes a dynamical scale of the gauge interaction
and we adopt a naive dimensional analysis
\cite{Luty}.

The effective K{\" a}hler potential may be expressed as
\beq
K_{{\rm eff}} = |S|^2 - \sum_{n=1} {k_n \o 16\pi^2 (n+1)^2 \L^{2n}} |\l S|^{2(n+1)},
\label{kahler}
\eeq
for $|\l S| \lsim \L$ with $k_n$ of order one.%
\footnote{We abuse the notation of $S$ for its lowest component.
}

On the other hand, the perturbative K{\" a}hler potential is given by
\beq
K_{{\rm eff}} = |S|^2 - A|\l S|^2 \ln \left| {\l S \o \mu} \right|^2 + \cdots,
\eeq
for $|\l S| \gg \L$, where $\mu$ is a renormalization point and
$A = N_C N_F / 16\pi^2$.
%

\section{Metastable Supersymmetry Breaking}
\label{sec:3}

For an illustrative purpose, we only keep one term
in the summation of Eq.(\ref{kahler}) and assume its $k_n$ to be positive.
The effective scalar potential for $|\l S| \lsim \L$ is then given by
\beq
V_{{\rm eff}} \simeq  {\l^2 \L^4 \o (16\pi^2)^2}  {N_F^2 \o N_C^2} 
\left( 1 + {k_n \l^2 \o 16\pi^2} \left| {\l S \o \L} \right|^{2n} \right)
\left| {\l S \o \L} \right|^{-a},
\eeq
where $a = 2(1-N_F/N_C)$.
For $N_F < N_C$, this gives a local minimum at
\beq
|\l S_{\rm min}| \simeq \L \left( {16\pi^2 a \o k_n \l^2 (2n-a)} \right)^{1 \o 2n}.
\eeq

The mass around the local minimum is
\beq
m_S \;\simeq\; \beta \frac{\lambda^2 \Lambda}{16 \pi^2},
\label{mass}
\eeq
with $\beta$ defined by
\beq
\beta \;\equiv\; \sqrt{na}   \left( {16\pi^2 a \o k_n \l^2 (2n-a)} \right)^{-\frac{a+2}{4n}} \frac{N_F}{ N_C}.
\eeq
The potential height at the minimum is
\beq
V_{{\rm eff}}(S_{\rm min}) \;\simeq\;\eta \frac{\lambda^2 \Lambda^4}{(16\pi^2)^2}\simeq 3 m_{3/2}^2 M_P^2,
\label{cc}
\eeq
with
\beq
\eta \;\equiv\; \lrf{2n}{2n-a} \left( {16\pi^2 a \o k_n \l^2 (2n-a)} \right)^{-{a \o 2n}} \frac{N_F^2}{N_C^2},
\eeq
where $M_P \simeq 2.4 \times 10^{18}$\,GeV is the reduced Planck mass.
Here we have used a fact that the potential height at the local minimum is 
related to the gravitino mass $m_{3/2}$
provided that the cosmological constant vanishes in supergravity.

On the other hand, the effective scalar potential for $|\l S| \gg \L$ is obtained as
\beq
V_{{\rm eff}} \simeq {\l^2 \L^4 \o (16\pi^2)^2} {N_F^2 \o N_C^2}
\left( 1 + A  \l^2 \ln \left| {\l S \o \mu} \right|^2 \right) \left| {\l S \o \L} \right|^{-a}.
\eeq
For $a$ of order one, this is monotonically decreasing within the perturbative validity.
We see that there are runaway directions with 
$|S| \rightarrow \infty$ (as well as directions $S=0$
with $Q_i {\tilde Q}_i$ running away for $N_F \neq 1$).
The tunneling into such directions
from the metastable state at $S=S_{\rm min}$ may well
be sufficiently suppressed
for small $\l$, which we assume in the following analyses.

\section{Polonyi Cosmology}
\label{sec:4}

Let us here explain the cosmological problem of the the Polonyi field $S$.
In the gravity mediation, $S$ must be a singlet under any symmetries
in order to generate SUSY standard-model gaugino masses. Since a singlet does not have
any special point in its field space, there is no reason to expect that the initial position 
$S_{\rm ini}$ happens to be very close to the SUSY-breaking (local) 
minimum. Generically, we expect that the deviation from the SUSY-breaking (local) minimum
is of $O(M_P)$.

After inflation, $S$ rolls down to the nearest minimum when the Hubble parameter 
becomes comparable to its mass. Suppose that the $S$ starts to oscillate about the 
SUSY-breaking minimum with an initial amplitude of $O(M_P)$. Then $S$  decays mainly
into a pair of the gravitinos
\footnote{
This Polonyi decay is not the exclusive source of the gravitino
production. In addition to the usual thermal production processes,
the gravitinos can be produced non-thermally by the inflaton decay.
In particular,
the singlet SUSY-breaking field can couple to the inflaton sector
so strongly that prohibitively large amount of gravitinos may be produced
\cite{Kawa}, though this depends on the details of the inflaton sector.
For instance, the chaotic inflation model with a $Z_2$ symmetry
causes no problem in this regard.}, 
whose decay produces energetic particles and 
significantly changes the light-element abundances~\footnote{
Stable gravitinos may result in overclosure of the universe. Moreover,
the next-to-lightest SUSY particle may significantly affect the light-element 
abundances, making the situation even worse.}.
Thus the abundance of the gravitinos produced by the $S$ decay will be
in conflict with BBN.
This is the notorious Polonyi problem
\footnote{See Ref.\cite{Polonyi}.
Several solutions have been proposed so far
(see Ref.\cite{Rand}).
}.

In the following, we point out
a possibility to avoid the Polonyi problem based on a class of 
the SUSY-breaking models presented in the previous section.

\subsection{Inflationary stage}

Let us first consider how the initial position $S_{\rm ini}$
of the $S$ field is determined. The K\"ahler potential $K$ generically contains a linear term of $S$,
\begin{equation}
      K\;=\; c\, M_P S + c^* M_P S^\dag + \cdots,
       \label{eq:linear}
\end{equation}
where $c$ is a numerical coefficient.
Although the effects of this term on the local minimum
$|S_{\rm min}| \ll M_P$ may be neglected due to Planck suppression in supergravity,
they possibly affect the behavior of $S$ during and after inflation.

The effective potential for $S$ after inflation before the reheating is given by%
\footnote{ It should be noted that $V_{{\rm eff}}(S)$ is absent until
$H$ becomes smaller than or equal to $\Lambda$. However, this does not change our conclusion.}
\bea
V(S) &\simeq& e^{K/M_P^2} (3 H^2 M_P^2) + V_{{\rm eff}}(S) \non\\
	  &\simeq& 3 H^2 \left( |S|^2 + c M_P S + c^* M_P S^\dag \right) 
	    	    + V_{{\rm eff}}(S).
\label{eq:V_s}		    
\eea

When the Hubble parameter during the inflation satisfies
$H_{\rm inf} \gg m_{S}$, the initial deviation $\Delta S$ is given by
\begin{equation}
       \Delta S \;\equiv\; |S_{\rm ini}-S_{\rm min}| \simeq |c| M_P,
       \label{sini_largeH}
\end{equation}
for $|S_{\rm ini}| > |S_{\rm min}|$. The $S$ remains at $S_{\rm ini}$ until it starts to oscillate when $H \simeq m_S$~\footnote{
The discussion here can be applied to a class of the $D$-term inflation models
in which the Hubble-induced mass of the $S$ field appears after inflation ends.
This is indeed the case in the known $D$-term inflation models.
}.

For a low scale inflation with $H_{\rm inf} \ll m_{S}$, the fate of the evolution of $S$ 
depends on the initial condition. If $S$ happens to be
close to the SUSY-breaking vacua, $|S| \lsim \Lambda/\lambda$, the typical
 deviation  is given by
\beq
\Delta S \;=\; |S_{\rm ini}-S_{\rm min}| \sim |c|\, \frac{H^2}{m_S^2} M_P,
\label{eq:sini-low}
\eeq
which is time-dependent. 
After inflation, $S$ gradually approaches $S_{\rm min}$ as the
Hubble parameter decreases, and no sizable coherent oscillations of $S$ are
induced~\cite{Linde:1996cx}.
On the other hand, if the initial $|S|$ is larger than $\Lambda/\lambda$,
it will fall into one of the SUSY minima during inflation.

For completeness let us also consider a case
that there is no Hubble-induced contribution to the Polonyi potential during and
after inflation. In this case the $S$ field acquires quantum fluctuations during inflation with an
amplitude given by $\delta S = H_{\rm inf}/2\pi$, which are not damped 
until $S$ starts oscillating.  Therefore, the inflationary scale must satisfy
$H_{\rm inf}/2\pi \lsim \Lambda/\lambda$ for the $S$ field in order not to fall into one of the SUSY minima.

\subsection{Reheating stage}

Suppose that the initial displacement from the local SUSY-breaking minimum is determined  in an anthropic
way. That is to say, there are many sub-universes where $S_{\rm ini}$ takes randomly
chosen values, and our observable universe is contained in one of them.
\footnote{When the initial displacement is given by (\ref{sini_largeH}), this may be the case if
$c$ takes different values in different sub-universes that are far apart from each other.}

If a value of $S_{\rm ini}$ in a sub-universe 
is such that the $S$ falls into one of the SUSY vacua in the end, we can simply
discard such a sub-universe, since no life like us will be formed.  Therefore, only the sub-universes with
the value of $S_{\rm ini}$ much smaller than $M_P$ are selected
anthropically, and the deviation from the local minimum is necessarily 
smaller than $\Lambda/\lambda$ in such sub-universes.

We first consider the inflation with $H_{\rm inf} \gg m_S$ in the following.
Since there is no anthropic argument for suppressing 
the initial deviation to be much smaller than $\Lambda/\lambda$,
we expect that the typical deviation leading to the habitable universe is 
of order $\Lambda/\lambda$.

Let us now estimate the cosmological abundance of $S$. The timing of the reheating
is crucial for determining the $S$ abundance. As we will see later, if the reheating occurs
before the $S$ starts to oscillate, too many gravitinos would  be produced by thermal scatterings.
Therefore we assume that  the reheating occurs after the $S$ starts to oscillate when the Hubble 
parameter becomes comparable
to its mass, $H \simeq m_S$. Then the abundance of $S$ is
\bea
\frac{\rho_S}{s} \simeq \left.\frac{\rho_{\rm inf}}{s}\right|_{\rm R} \left.\frac{\rho_S}{\rho_{\rm inf}}\right|_{\rm osc}
				\simeq \frac{3 T_R}{4} \cdot \frac{m_S^2\, \Delta S^2}{3 m_S^2 M_P^2}
				= \frac{T_R}{4} \frac{\Lambda^2}{\lambda^2 M_P^2} \lrfp{\Delta S}{\Lambda/\lambda}{2},
\label{rhos}				
\eea
where $\rho_S$ and $\rho_{\rm inf}$ denote the energy densities
of $S$ and the inflaton, $s$ the entropy density,
$T_R$ the reheating temperature, $m_S$ the mass of $S$, and
the subscripts ``R" and ``osc" mean that the variables are evaluated at the reheating and at the commencement of 
the oscillations, respectively.%

The condensate of $S$ will dominantly decay into a pair of the gravitinos. The abundance of the gravitinos produced
from the $S$ decay
is therefore related to the $S$ abundance as
\beq
Y_{3/2}^{(S)} \;\simeq\; \frac{2}{m_S} \frac{\rho_S}{s}.
\eeq
Eqs.\REF{mass} and \REF{cc} result in
\beq
Y_{3/2}^{(S)} \;\simeq\; 1 \times 10^{-17} \,\beta^{-1} \eta^{-\frac{1}{4}} \lambda^{-\frac{9}{2}} 
 \lrfp{|S_{\rm ini}|}{\Lambda/\lambda}{2}
 \lrfp{m_{3/2}}{1{\rm \,TeV}}{\frac{1}{2}} \lrf{T_R}{10^6{\rm \,GeV}}.
\label{Ys}
\eeq
Note that $Y_{3/2}^{(S)}$ gets enhanced both for a heavier gravitino mass and for 
a smaller $\lambda$.

In addition to the production from the $S$ decay, the gravitinos are produced by
particle scatterings in thermal plasma. The abundance is approximately given by
\beq
Y_{3/2}^{(th)} \;\simeq\; 2 \times 10^{-16}  \lrf{T_R}{10^6{\rm \,GeV}},
\label{eq:th-grav}
\eeq
where we have neglected the contributions from the longitudinal mode.
Thus, unless $\lambda$ is much smaller than unity,
 the gravitinos from the $S$ decay tends to be smaller than or comparable  to
 the thermally produced gravitinos. In this respect our scenario makes the Polonyi problem 
less severe than the cosmological problem of the thermally produced gravitinos.

The gravitino abundance is tightly constrained by BBN.
According to the latest analysis~\cite{Kawasaki:2004yh},
the constraint is given by
\begin{eqnarray}
\label{eq:unstable-Y1}
   Y_{3/2}  & \lsim & \left\{\begin{array}{lcl}
   ~1\times 10^{-16} - 6\times 10^{-16}
   &{\rm for}    &  m_{3/2} \simeq 0.1 - 0.2~{\rm TeV} \\[0.8em]
   ~4\times 10^{-17} - 6\times 10^{-16}
   &{\rm for}    &  m_{3/2} \simeq 0.2 - 2~{\rm TeV} \\[0.8em]
   ~ 7 \times 10^{-17} - 2\times 10^{-14} 
   &{\rm for}    & m_{3/2} \simeq 2 - 10~{\rm TeV}
\end{array}\right..
\label{YBBN}
\end{eqnarray}
We require that the reheating temperature should be low enough for the thermally produced
gravitinos (\ref{eq:th-grav}) to be compatible with the BBN bounds. 
Then we can see that the abundance of the gravitinos produced by the $S$ decay can 
satisfy the BBN bound for $|S_{\rm ini}| \lsim \Lambda/\lambda$, unless $\lambda$ is much 
smaller than unity.
Note also that the gravitino abundance (\ref{eq:th-grav}) will be in conflict with the BBN
bounds if the reheating occurs before the $S$ starts oscillating, since $T_R$ would
exceed $10^{10}$\,GeV. This justifies our assumption in deriving (\ref{rhos}).

It may be worth mentioning that the abundance (\ref{Ys}) is
relatively close to the upper bound. For some choices of the ${\cal O}(1)$ parameters,
the gravitino abundance from $S$ decay may become inconsistent with the
BBN constraints for $T_R \gsim 10^6$\,GeV.%
\footnote{
This is especially the case if $T_R$ is close to $10^6$\,GeV, which is the lower bound coming from the non-thermal 
leptogenesis to work. Thus the Polonyi problem with the non-thermal leptogenesis scenario 
is marginally consistent with BBN.}
For a lower reheating temperature or a heavier gravitino mass, 
the gravitino abundance from the $S$ decay is
well below the BBN constraints, and the Polonyi problem is indeed absent.

Lastly we comment on a case of the low-scale inflation with $H_{\rm inf} \ll m_S$.
As mentioned before, the cosmological  abundance of $S$ is negligibly small since no sizable 
coherent oscillations are produced~\cite{Linde:1996cx}. Thus the Polonyi problem
is absent in this case. On the other hand, the above discussion can be applied to
a case that there is no Hubble-induced contribution to the $S$ field during and after inflation,
whereas a non-trivial constraint on the inflationary scale, $H_{\rm inf}/2\pi \ll \Lambda/\lambda$,
must be imposed.

\section{Conclusion}

We proposed a class of dynamical models
with SUSY-breaking metastable vacua
which are surrounded by SUSY ones.

The Polonyi problem in gravity mediation of SUSY breaking
can be avoided when our SUSY-breaking vacuum is surrounded
by such SUSY vacua. Then, the initial value
of the Polonyi field $S$ should be chosen not too far away from our
vacuum, since otherwise, the $S$ would roll down to the SUSY vacua.

We emphasize that the concrete models are mere examples
and the above circumstances might be generic enough.
The absence of the Polonyi problem may be a consequence
of the fact that our SUSY-breaking vacuum is metastable
if gravity mediation is realized in Nature.

\section*{Acknowledgements}

This work was supported by the Grant-in-Aid for Yukawa International
Program for Quark-Hadron Sciences, the Grant-in-Aid
for the Global COE Program "The Next Generation of Physics,
Spun from Universality and Emergence", and
World Premier International Research Center Initiative
(WPI Initiative), MEXT, Japan.

\end{document}